\documentclass{aa}  
\usepackage{graphicx}
\usepackage{txfonts}

\newcommand{\h}{$^{\rm h}$}
\newcommand{\m}{$^{\rm m}$}
\newcommand{\s}{$^{\rm s}$}

\newcommand{\ha}{\rm H$\alpha$}
\newcommand{\hbeta}{\rm H$\beta$}

\newcommand{\hnii}{{\rm H}$\alpha+[$\ion{N}{ii}$]$}
\newcommand{\nii}{$[$\ion{N}{ii}$]$}
\newcommand{\sii}{$[$\ion{S}{ii}$]$}
\newcommand{\oi}{$[$\ion{O}{i}$]$}

\newcommand{\oiii}{$[$\ion{O}{iii}$]$}
\newcommand{\snr}{\rm supernova remnant}

\newcommand{\flux}{$10^{-17}$ erg s$^{-1}$ cm$^{-2}$ arcsec$^{-2}$}
\newcommand{\dens}{\rm cm$^{-3}$}
\newcommand{\vel}{\rm km s$^{-1}$}
\newcommand{\sulfur}{[S~{\sc ii}]}
\newcommand{\nitrogen}{[N~{\sc ii}]}

\newcommand{\siirat}{$[$\ion{S}{ii}$]\lambda\lambda\ 6716/6731$} 
\newcommand{\HNII}{{\rm H}$\alpha+$[N {\sc ii}]~6548~\&~6584~\AA}  
\newcommand{\OIII}{$[$\ion{O}{iii}$]$~5007~\AA}
\newcommand{\SII}{$[$\ion{S}{ii}$]$~6716~\&~6731~\AA}

\begin{document}
\title{Discovery of optical emission from the supernova remnant G
32.8$-$0.1 (Kes 78)}


\author{P. Boumis\inst{1}
\and E. M. Xilouris\inst{1}
\and J. Alikakos\inst{1,2}
\and P. E. Christopoulou\inst{2}
\and  F. Mavromatakis\inst{3}
\and A.C. Katsiyannis\inst{1}
\and C. D. Goudis\inst{1,2}
}

\offprints{P. Boumis}

\authorrunning{P. Boumis et al.}
\titlerunning{Discovery of optical emission from the SNR G 32.8$-$0.1}

\institute{Institute of Astronomy \& Astrophysics, National
Observatory of Athens, I. Metaxa \& V. Paulou, P. Penteli, GR-15236
Athens, Greece.\\
\email{[ptb;xilouris;johnal;thk;cgoudis]@astro.noa.gr}
\and Astronomical Laboratory, Department of Physics, University of
Patras, 26500 Rio-Patras, Greece.\\
\email{pechris@upatras.physics.gr}
\and Technological Education Institute of Crete, Department of
Sciences, P.O. Box 1939, GR-710 04 Heraklion, Crete, Greece.\\
\email{fotis@stef.teiher.gr} }

\date{Received / accepted }

\abstract{Deep optical CCD images of the supernova remnant G
32.8$-$0.1 were obtained where filamentary and diffuse emission was
discovered. The images were acquired in the emission lines of \hnii\
and \sulfur. Filamentary and diffuse structures are detected in most
areas of the remnant, while no significant \oiii\ emission is
present. The flux--calibrated images suggest that the optical emission
originates from shock-heated gas since the \sulfur/\ha\ ratio is
greater than 1.2.  The Spitzer images at 8$\mu$m and 24$\mu$m show a
few filamentary structures to be correlated with the optical
filaments, while the radio emission at 1.4 GHz in the same area is
found to be very well correlated with the brightest optical
filaments. Furthermore, the results from deep long--slit spectra also
support the origin of the emission to be from shock--heated gas
(\sulfur/\ha\ $>$ 1.5). The absence of \oiii\ emission indicates slow
shocks velocities into the interstellar $``$clouds'' ($\leq$100 \vel),
while the \siirat\ ratio indicates electron densities up to $\sim$200
cm$^{-3}$. Finally, the \ha\ emission is measured to lie between 1.8
to 4.6 $\times$ \flux, while from VGPS H{\sc i} images a distance to
the SNR is estimated to be between 6 to 8.5 kpc.
}

\maketitle
%

\section{Introduction}

Supernova remnants (SNRs) are important components of a galaxy since
during the supernova ejecta they release huge amounts of energy and in
subsequent phases heavy elements are mixed into the interstellar
medium (ISM). Hence, they play an important role to understand the
supernovae mechanism, the interstellar medium (ISM) and their
interaction. The majority of the SNRs have been discovered in radio
wavelengths from their non-thermal synchrotron emission and only a few 
of them have been detected in soft X-rays and in the optical (Green
\cite{gre06}). Optical observations (i.e. Boumis et al. \cite{bou05};
\cite{bou08}) offer an important tool for the study of the interaction
of the shock wave with dense concentrations of gas found in the ISM.

The Galactic SNR G 32.8--0.1 (Kes78) was discovered in a 408 MHz radio
continuum survey by Kesteven (\cite{kes68}), where it appears as a
small source (20\arcmin $\times$ 10\arcmin\ in diameter). Several
radio observations were perfromed eversince at 30.9, 330, 408, 430,
2700, 5000, 10600 MHz by Kassim (\cite{kas89}), Kassim (\cite{kas92}),
Caswell et al. (\cite{cas75}), Dickel \& Denoyer (\cite{dic75}),
Velusamu \& Kundu (\cite{vel74}), Caswell et al. (\cite{cas75}) and
Beckel \& Kundu (\cite{bec75}) showing its non--thermal emission,
while a spectral index between 0.5 and 0.78 was
established. Observations in the hydroxyl molecule (OH) at 1720 MHz
(Koralesky et al. \cite{kor98}) showed that the emission results from
masers that are shock--excited due to the interaction of the SNR and
an adjacent molecular cloud. They found a magnetic field of 1.5 $\pm$
0.3 mG and determined a kinematic distance to the SNR at 7.1
kpc. Allakhverdiyev et al. (\cite{all83}) and Case \& Bhattacharya
(\cite{cas98}) using the radio surface brightness -- diameter
relationship ($\Sigma$--D) calculated a distance at 7.1 kpc and 6.3
kpc respectively. Neutral hydrogen observations at 21 cm (Gosachinskii
\& Khersonskii \cite{gos85}) resulted to a distance of 9 kpc, an age
of 1.2 $\times 10^{5}$~yr and an explosion energy of 5.3 $\times
10^{50}$~erg for the SNR. In radio surveys of the surrounding region
at 1.4 GHz (Manchester et al. \cite{man85}), no pulsar was found to be
associated with G 32.8--0.1 up to $\sim$1 mJy of period $\ge$ 10 ms.
IRAS observations (Saken et al. \cite{sak92}) showed a barely resolved
IR shell to be coincident with the SNR's radio emission.  Spitzer
Space Telescope observed the region of interest as part of a large
survey to map a large part of the Galactic plane. In a recent study
(Reach et al. \cite{rea06}), the radio emission of Kes 78 was compared
with the infrared emission of this part of the sky and the comparison
was described as ``confused'' since no clear correlation was evident.

In this paper we report the optical detection of G 32.8--0.1 and
present flux calibrated images in major optical
emission lines. Deep long slit spectroscopy was also performed in a
number of selected areas of interest. In Sect. 2, information about
the observations and data reduction is given, while the results of the
imaging and spectroscopic observations are presented in Sect. 3 and
4. In Sect. 5, we compare our data with observations at other
wavelengths, in Sect. 6 we discuss the optical properties of the SNR,
while in Sect. 7 we summarize the results of this work.

\section{Observations}
A summary and log of our observations are given in
Table~\ref{table1}. In the sections below, we describe these 
observations in detail.

\subsection{Imaging}
\subsubsection{Wide--field imagery}

\begin{figure*}
\centering
\includegraphics[]{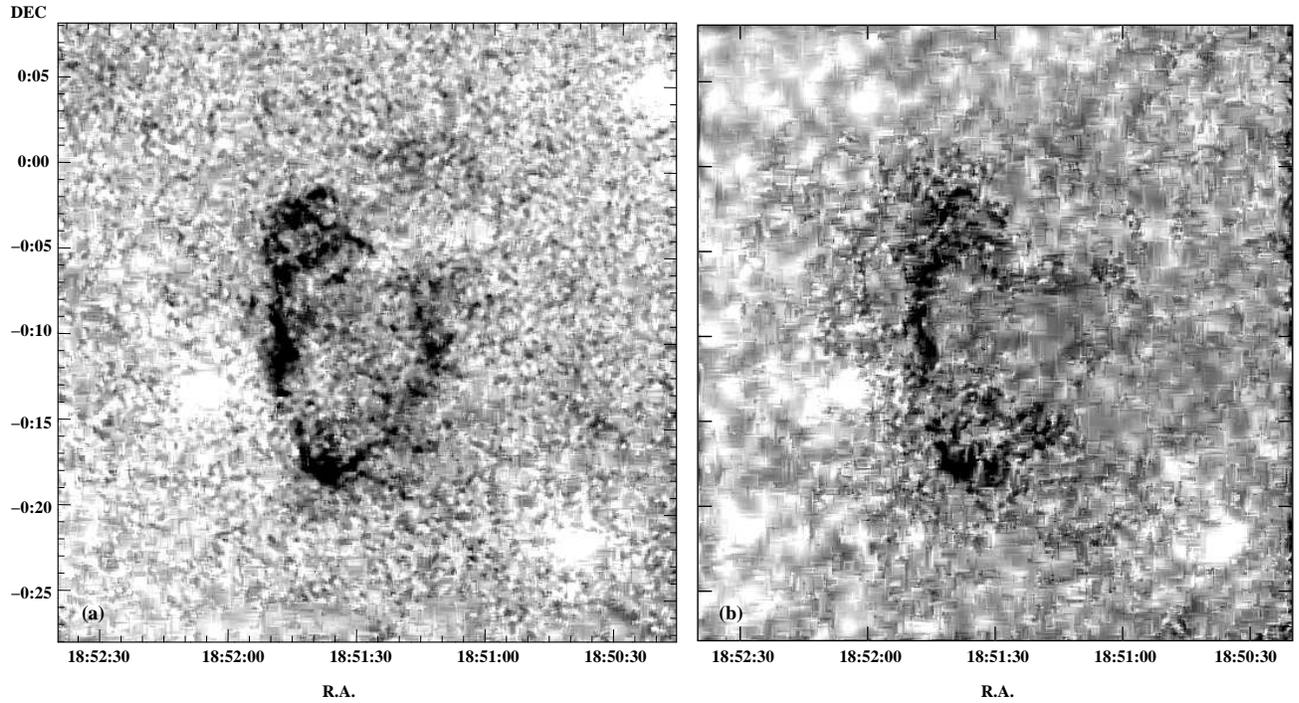}
\caption{The G 32.8$-$0.1 in (a) the H$\alpha$ $+$ [N {\sc ii}] and
(b) the \sii\ filter. Shadings run linearly from 0 to 40$\times
10^{-17}$~erg s$^{-1}$ cm$^{-2}$ arcsec$^{-2}$~and 0 to 25$\times
10^{-17}$~erg s$^{-1}$ cm$^{-2}$ arcsec$^{-2}$~ in (a) and (b),
respectively. The images have been smoothed to suppress the residuals
from the imperfect continuum subtraction.}
\label{fig1}
\end{figure*}

The wide--field imagery of G 32.8--0.1 was obtained with the 0.3 m
Schmidt--Cassegrain (f/3.2) telescope at Skinakas Observatory, Crete,
Greece in June 7--9, 11 and August 26, 2005. A 1024 $\times$ 1024
Thomson CCD was used which has a pixel size of 19 $\mu$m resulting in
a 70\arcmin\ $\times$ 70\arcmin\ field of view and an image scale of
4\arcsec\ per pixel. The area of the remnant was observed with the
\hnii, [S~{\sc ii}], and [O~{\sc iii}] filters. The exposure time was
set to 2400 s for each exposure and 180 s for the continuum red
and blue filters. The corresponding continuum images were subtracted
from those containing the emission lines to eliminate the confusing
star field (see Boumis et al. \cite{bou02}, for details of this
technique).  The continuum--subtracted images of the \hnii\ and
[S~{\sc ii}] emission lines are shown in Fig.~\ref{fig1}(a) and
\ref{fig1}(b), respectively.

The IRAF and MIDAS packages were used for the data reduction. All
frames were bias--subtracted and flatfield--corrected using a series
of well exposed twilight frames. A smoothing of 5 $\times$ 5
pixels median box was applied to all images while the stars were
removed by appropriately scaling and subtracting the continuum images.
The spectrophotometric standards stars HR5501, HR7596, HR7950, HR8634,
and HR9087 (Hamuy et al. \cite{ham92}) were used for the absolute flux
calibration. The astrometric solution for all data frames was
calculated using the Hubble Space Telescope (HST) Guide Star Catalogue
(Lasker et al. \cite{las99}). All the equatorial coordinates quoted in
this work refer to epoch 2000.

For all images we calculated median smoothing over a 5x5 neighborhood
and replace the original value by the median value only using MIDAS
command "filter/median". A sentence was added in the text to make it
clear.

\subsubsection{High--resolution imagery}

\begin{figure*}
\centering
\includegraphics[bb=100 50 753 587]{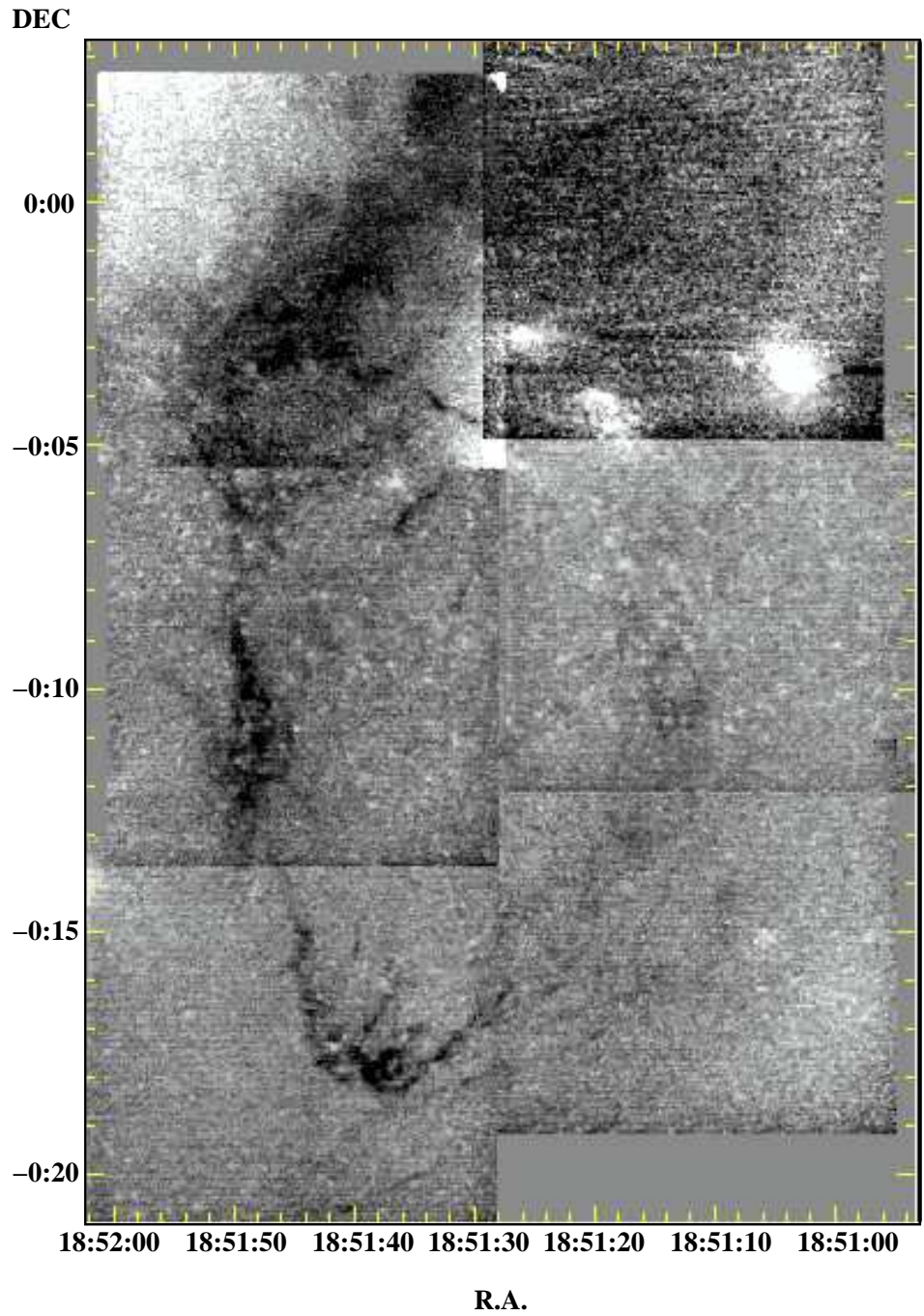}
\caption{The continuum--subtracted mosaic of G 32.8$-$0.1 taken with
the 1.3--m telescope in the light of H$\alpha$$+$[N {\sc ii}]. The
image has been smoothed to suppress the residuals from the imperfect
continuum subtraction.}
\label{fig2}
\end{figure*}

Optical images at higher angular resolution of G 32.8$-$0.1 were also
obtained with the 1.3 m (f/7.7) Ritchey-- Cretien telescope at
Skinakas Observatory in July 4--7 and 8--10, 2007, using the
H$\alpha$+[N {\sc ii}] interference filter. The detector was a 1024
$\times$ 1024 SITe CCD with a field of view of 8.5 $\times$ 8.5
arcmin$^{2}$~and an image scale of 0.5\arcsec\ per pixel. Six
exposures were taken through the H$\alpha$+[N {\sc ii}] filter, each
of 2400 s and six corresponding exposures in the continuum, each of
180 s. After the continuum subtraction, all fields were projected to a
common origin on the sky and were subsequently combined to create the
final mosaic in \hnii. Note that the upper right field was
observed close to the morning twilight, significantly reducing  its
signal--to--noise in comparison with the surrounding fields. During
the observations, the ``seeing'' varied between 0.8\arcsec and
1.5\arcsec, while the full width at half maximum (FWHM) of the star
images was between 1.1\arcsec and 2.1\arcsec. The
continuum--subtracted mosaic is shown in Fig. \ref{fig2}.

\subsection{Spectroscopy}
Low dispersion long--slit spectra were obtained with the 1.3 m
telescope at Skinakas Observatory in June 4 and 5 and September 7,
2005. The exposure time was 3900 s. The 1300 line mm $^{-1}$ ~grating
was used with the 2000 $\times$ 800 (13 $\mu$m) SITe CCD covering the
range 4750\AA\ -- 6815\AA. The 1 \AA\ pixel$^{-1}$ and the
dispersion of 1300 line mm $^{-1}$~result in a spectral resolution of
$\sim$8 and $\sim$11 \AA\ in the red and blue wavelengths,
respectively. The slit has a width of 7\farcs7 and length of 7\farcm9
and in all cases was oriented in the south--north direction. The
coordinates of the slit centers of each spectrum are given in
Table~\ref{table1}. For the absolute flux calibration the
spectrophotometric standard stars HR4468, HR5501, HR7596,HR7950,
HR8634 and HR9087 were used. The data reduction was performed by using
the IRAF package.

%
%
\section{The \hnii, \sii\ and \oiii\ emission line images}

Optical emission is detected for the first time from this remnant. The
most interesting regions lie in the south, east and north, where
bright filamentary and faint diffuse structures are present (between $\alpha
\simeq$ 18\h51\m35\s, $\delta \simeq$ --00\degr17\arcmin30\arcsec;
$\alpha \simeq$ 18\h51\m51\s, $\delta \simeq$
--00\degr08\arcmin40\arcsec and $\alpha \simeq$ 18\h51\m41\s, $\delta
\simeq$ --00\degr01\arcmin20\arcsec), which all are very well
correlated with the radio emission. The bright \hnii\ filaments cover
most of the emission found in radio wavelengths and extend for
$\sim$16\arcmin\ from south to north through the east part of the
SNR. In contrast to above results, diffuse emission was mainly
detected in the west with only one bright filament between $\alpha
\simeq$ 18\h51\m09\s, $\delta \simeq$ --00\degr09\arcmin45\arcsec and
$\alpha \simeq$ 18\h51\m14\s, $\delta \simeq$
--00\degr10\arcmin48\arcsec\ ($\sim$1\arcmin\ long). No significant
emission was found in the image of the \oiii\ medium ionization line
and therefore it is not shown here. The morphology of the \sii\ image
is generally similar, though not as bright as, to that of the \hnii\
image. We detected \sii\ emission where most of the \hnii\ emission was
found with filamentary bright structures in the south and east areas,
while diffuse emission characterizes the rest of the remnant's
emission.

The flux calibrated images of \hnii\ and \sii\ provide a
first indication to the nature of the observed emission (see
Table~\ref{table2}). A study of these images shows that all parts of
the optical remnant originate from shock heated gas since we estimate
ratios \sii/\ha $\geq$1.2 consistent with the
spectral measurements. In particular, the eastern and southern areas
show \sii/\ha $\sim$1.5 and 1.7 respectively, while the northern area
shows \sii/\ha $\sim$1.3. Only the western area does not show strong
\sii\ emission, hence only upper limits are given.

\section{The optical spectra from G 32.8$-$0.1}
Deep low--resolution spectra were taken on the brightest optical
filaments in the eastern and southern parts of the SNR (their exact
position are given in Table~\ref{table1}). In Table~\ref{table3}, we
present the relative line fluxes taken from two different apertures
along the slit. In particular, apertures I and II have an offset (see
Table~\ref{table1}) north or south of the slit center 
because they are free of field stars and at the same time they include
sufficient line emission to alloow for an accurate determination of the
line fluxes. The background extraction aperture was taken
towards the northern or the southern ends of the slit depending on
the filament's position within the slit.  

\begin{figure*}
\centering
\includegraphics{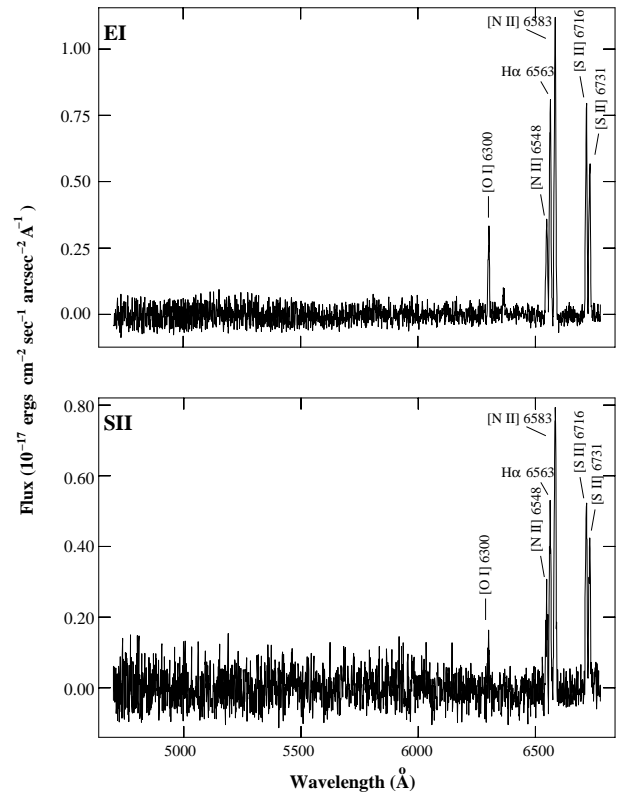}
\caption{Long--slit spectra from the east (EI) and south (SII) areas
of G 32.8$-$0.1.}
\label{fig3}
\end{figure*}

The measured line fluxes indicate emission from shock--heated gas,
since \sii/\ha\ $>$1.5. Furthermore, the \nii/\ha\ ratio, which takes
values between 1.3 and 2.2 (see Table~\ref{table3}), falls well inside
the range expected for an SNR (Fesen et al. \cite{fes85}). The
signal--to--noise ratios do not include calibration errors, which are
less than 10 percent. Typical spectra from the east (E1) and south
(SII) are shown in Fig.~\ref{fig3}. The absolute \ha\ flux covers a
range of values from 1.8 to 4.6 $\times$ \flux. The \siirat\ ratio
that was measured between 1.3 and 1.5, indicates electron densities
below 150 cm$^{-3}$~ (Osterbrock \& Ferland \cite{ost06}). However,
taking into account the statistical errors on the sulfur lines, it is
found that electron densities up to 200 cm$^{-3}$ are allowed (Shaw \&
Dufour \cite{sha95}). \oiii\ emission was not detected, while the very
weak \hbeta\ emission and the absence of the \oiii\ suggest
significant interstellar attenuation of the optical emission. The
absence of \oiii\ emission can also be explained by slow shocks
propagating into the interstellar clouds ($\leq$ 100 \vel; Hartigan et
al. \cite{har87}) since higher velocities shocks should produce
detectable \oiii\ emission.

\section{Observations at other wavelengths}
\begin{figure*}
\centering
\includegraphics[bb=120 30 766 700]{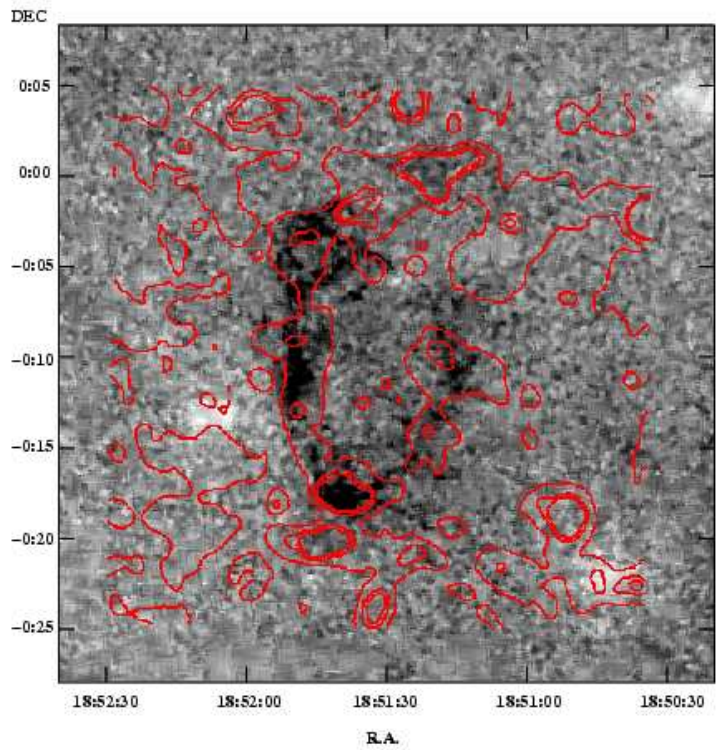}
\caption{The correlation between the H$\alpha +$[N {\sc ii}] emission,
and the radio emission at 1.4 GHz (NVSS -- solid red line). The 1.4
GHz radio contours scale linearly from 6$\times 10^{-3}$ Jy/beam, to
0.02 Jy/beam.}
\label{fig4}
\end{figure*}

The optical emission matches the radio emission of G 32.8$-$0.1 very
well at 1.4 GHz, suggesting their association (Fig.~\ref{fig4}). The
observed filaments are located close to the outer edge of the radio
contours but the low resolution of the radio images does not allow us
to determine the relative position of the filament with respect to the
shock front. We investigate possible correlations of the optical
emission with infrared emission.

\begin{figure*}
\centering
\includegraphics[bb=30 80 801 677]{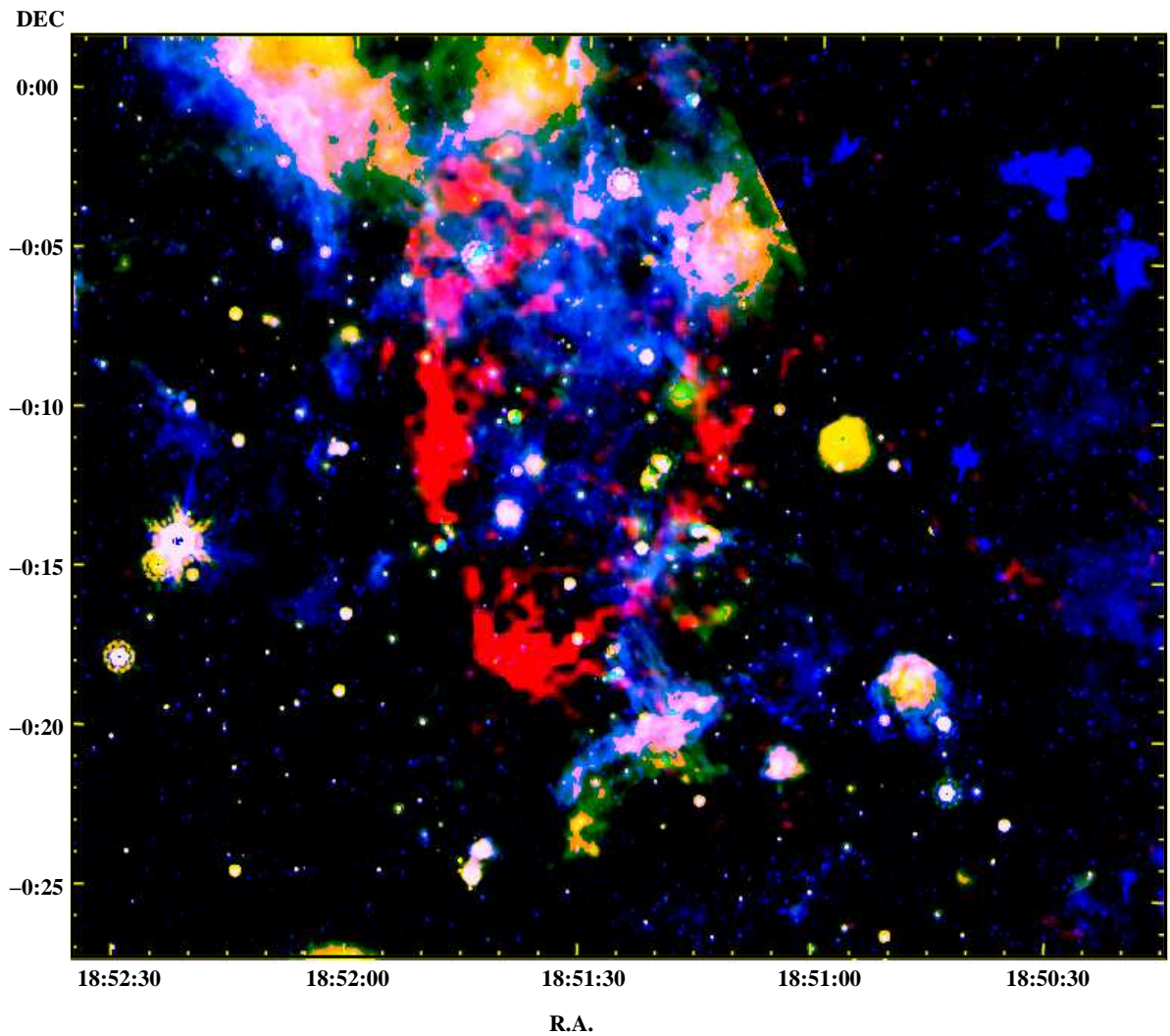}
\caption{The comparison between the H$\alpha +$[N {\sc ii}] emission
(red), and the infrared emission at Spitzer $8 \mu$ (blue)
and $24 \mu$ (green).  The $24 \mu m$ emission shadings run
logarithmic from 45 to 105 MJy/sr and the $8 \mu m$ emission from 66
to 240 MJy/sr.}
\label{fig5}
\end{figure*}

The region of interest has been observed with the Spitzer Space
Telescope in both $8 \mu m$ and $24 \mu m$ bands as part of the large
surveys GLIMPSE and MIPSGAL (Benjamin et al. \cite{ben03} and Carey et
al. \cite{car05}, respectively) along the Galactic plane. The superb
resolution of Spitzer in these wavelengths allowed us to perform a
very careful comparison between optical and infrared emission and look
for possible correlations.  In Fig.~\ref{fig5}, we present a composite
image of the \ha\ optical emission (red), the $24~\mu m$ emission
(green) and the $8~\mu m$ emission (blue).  It is evident that
although there is quite a lot of structure and confusion in the
background due to neighboring emission regions (as already pointed out
by Reach et al. \cite{rea06}), there are regions (e.g. in the North,
North--West and South) where the optical emission overlaps with
infrared (mostly with the $8~\mu m$) emission. We also argue that the
discontinuity of the shape of the remnant in the North--West part may
be due to heavy obscuration due to dust clouds seen in the
infrared. 

\begin{figure*}
\centering
\includegraphics[bb=-20 0 548 615]{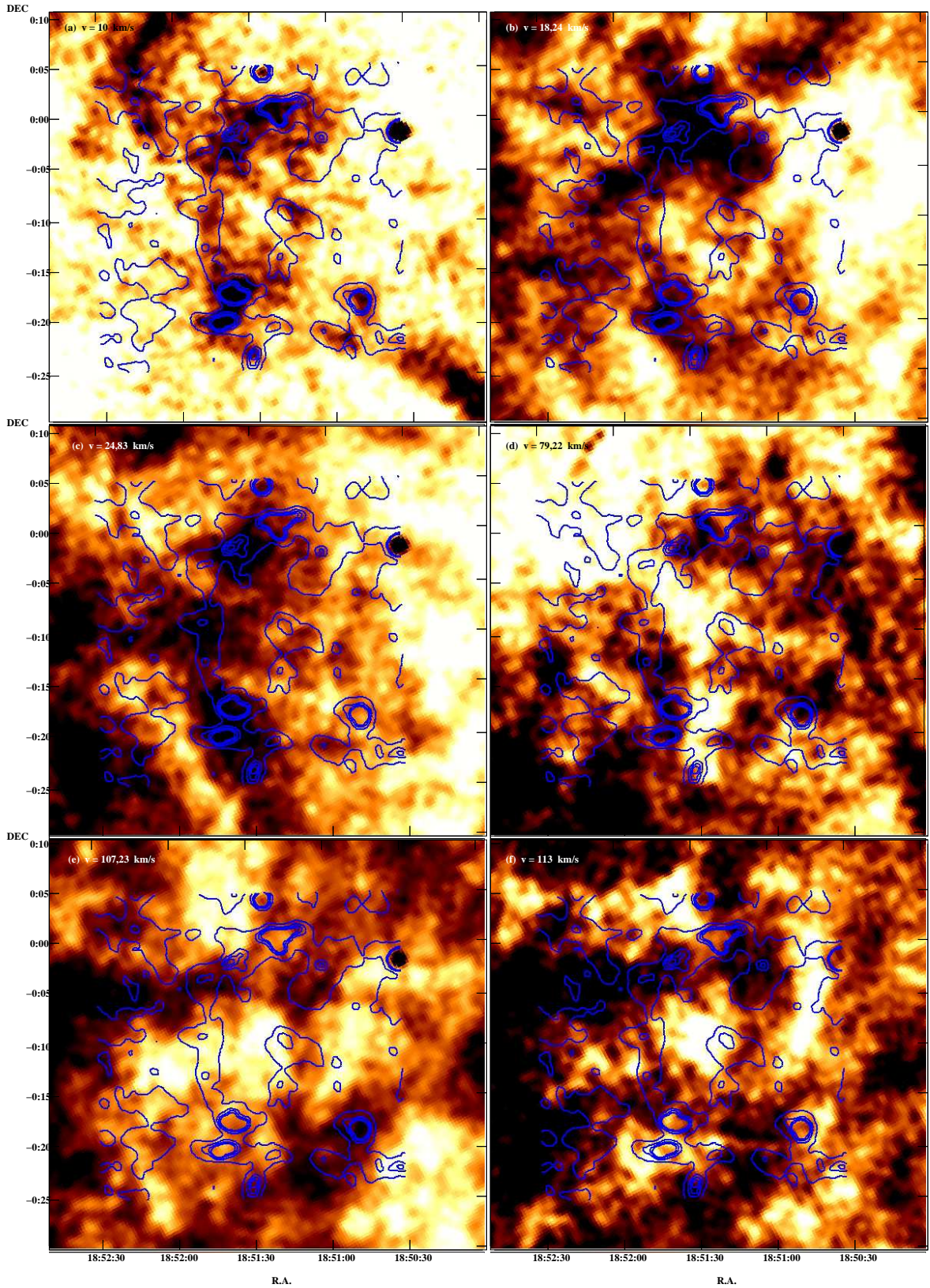}
\caption{VGPS images of the 21 cm line correlated with the radio
emission of G 32.8$-$0.1. The LSR velocities are indicated on the top
left of each panel. The VGPS images scale linearly from 10 to 130 K
for panels (a)--(c) and from 0 to 120 K for panels (d)--(f). The 1.4
GHz radio contours (blue line) scale linearly from 6$\times 10^{-3}$
Jy/beam, to 0.02 Jy/beam.}
\label{fig6}
\end{figure*}

New H{\sc i} kinematics data are available from the VLA Galactic Plane
Survey (VGPS; Stil et al. \cite{sti06}). The distribution of the
atomic gas was examined in detail by searching radial velocities in
the range from $-$113 to 165 \vel\ and looking for signs of
interaction between the expanding shock fronts of the SNR and the
surrounding medium. In Fig.~\ref{fig6}, we present velocities from six
different channels and radio contours on top of the H{\sc i} emission.
The H{\sc i} images in the range of 10 to 25 \vel\ (Fig. 6(a)--(c))
are interesting because intense H{\sc i} emission is present almost
all over the SNR's radio boundaries. If the H{\sc i} emission is due
to the SNR's expanding envelope then using the Galactic rotation curve
model and recent measurements of its parameters (R$_{0}=7.6$ kpc,
Eisenhauer at al. \cite{eis05}; V$_{0}=220$ \vel\ Feast \& Whitelock
\cite{fea97}) we can estimate distances in the range of 0.8 to 2 kpc.
At velocities greater than 80 \vel\ (Fig. 6(d)--(e)) and especially at
113 \vel\ (Fig. 6(f)) features all around the remnant show signs of
interaction of the SNR with the H{\sc i} emission. In that case, the
distances are estimated to be in the range of 6 to 8.5 kpc.

ROSAT All--sky survey data were also examined, but no
significant X--ray emission was detected.

%

\section{Discussion}
The \snr\ G 32.8--0.1 shows up as an almost complete shell in the
radio band, its optical emission marginally correlates with the
infrared emission and no X--ray emission has been detected so far. The
absence of soft X--ray emission may indicate a low shock temperature
and/or a low density of the local interstellar medium. The \oiii\ flux
production depends mainly upon the shock velocity and the ionization
state of the preshocked gas. Therefore, as mentioned in Sect. 3.2, the
absence of \oiii\ emission may be explained by slow shocks propagating
into the ISM. The \hnii\ image best describes the newly detected
structures. Sulfur line emission is also detected and generally
appears less filamentary and more diffuse than in the \hnii\ image
with their position and shape to be in agreement with that of the
\hnii. The absence of \oiii\ emission does not allow us to determine
whether slow shocks travel into ionized gas or whether faster shocks
travel into neutral gas (Cox \& Raymond 1985) but we can exclude
moderate or fast shocks overtaking ionized gas. The presence of [O
{\sc i}] 6300 \AA\ line emission is also consistent with the emission
originating from shock heated gas. Both the calibrated images and the
long--slit spectra suggest that the detected emission results from
shock heated gas since the \sii/\ha\ ratio exceeds the empirical SNR
criterion value of 0.4--0.5, while the measured \nii/\ha\ ratio also
confirms this result.
\par

The interstellar extinction c cannot be accurately determined due to
the low significance of the \hbeta\ flux. However, using the \hbeta\
upper limits (see Table~\ref{table3}), the lower limits on c(\hbeta)
are calculated to 2.2 and 2.7 or A$_{\rm V}$ of 4.51 and 5.53 for the
areas in the South and the East, respectively. Both limits derived
from our spectra, suggest an area of high interstellar extinction as
it is expected for an object being at the galactic plane.

Estimated values of N$_{\rm H} \sim 1.9 \times 10^{22}$~cm$^{-2}$ and
N$_{\rm H} \sim 1.5 \times 10^{22}$~cm$^{-2}$ are given by Dickey \&
Lockman (\cite{dic90}) and Kalberla et al. (\cite{kal05})
respectively, for the column density in the direction of G
32.8$-$0.1. Using the relation of Ryter et al. (\cite{ryt75}), we obtain
an N$_{{\rm H}} >$ $9.9 \times 10^{21}~{\rm cm}^{-2}$~and $1.2 \times
10^{22}~{\rm cm}^{-2}$~for the c limits calculated
from our spectra, respectively. Both values are consistent with the
estimated galactic N$_{\rm H}$~considering the uncertainties involved.

We have also determined the electron density measuring the density
sensitive line ratio of \siirat. The measured densities lie below
200 \dens. Assuming that the temperature is close to 10$^{4}$ K, it is
possible to estimate basic SNR parameters. The remnant under
investigation has not been studied in detail hence the current stage
of its evolution is unknown. Assuming that the remnant is still in the
adiabatic phase of its evolution the preshock cloud density n$_{\rm
c}$ can be measured by using the relationship (Dopita \cite{dop79})

\begin{equation}
{\rm n_{[SII]} \simeq\ 45\ n_c V_{\rm s}^2}~{\rm cm^{-3}},
\end{equation}

where ${\rm n_{[SII]}}$ is the electron density derived from the
sulfur line ratio and V$_{\rm s}$ is the shock velocity into the
clouds in units of 100 \vel. Furthermore, the blast wave energy can be
expressed in terms of the cloud parameters by using the equation given
by McKee \& Cowie (\cite{mck75})

\begin{equation}
{\rm E_{51}} = 2 \times 10^{-5} \beta^{-1} 
{\rm n_c}\ V_{\rm s}^2 \ 
{\rm r_{s}}^3 \ \ {\rm erg}.
\end{equation}

The factor $\beta$ is approximately equal to 1 at the blast wave
shock, ${\rm E_{51}}$ is the explosion energy in units of 10$^{51}$
erg and {\rm r$_{\rm s}$} the radius of the remnant in pc. By using
the upper limit on the electron density of 200 \dens, which was
derived from our spectra, we obtain from Eq. (1) that ${\rm n_c}
V_{\rm s}^2 < 4.4$. Then Eq. (2) becomes ${\rm E_{51}} < 1.1 \times
10^{-3}~{\rm D_{1 kpc}^3}$, where ${\rm D_{1 kpc}}$~the distance to
the remnant in units of 1 kpc.
\par
Since, there are no other measurements of the interstellar density
n$_{0}$, values of 0.1 and 1.0 will be examined. Following the result
of Eq. (2) and assuming the value of 0.53 (Gosachinskii \& Khersonskii
1985) for the supernova explosion energy (E$_{51}$), we find that the
remnant may lie at distance greater than 7.9 kpc. Then, the higher
interstellar density of $\sim$1 cm$^{-3}$~suggests that the column
density is greater than $7.9 \times 10^{21}~{\rm cm}^{-2}$, while for
n$_{0} \approx 0.1~{\rm cm}^{-3}$~it becomes greater than $7.9 \times
10^{20}~{\rm cm}^{-2}$. Combining the previous results and assuming
that the column density is found in the range of $1.5 - 1.9 \times
10^{22}~{\rm cm}^{-2}$, then the higher interstellar density seems to
be more probable.  The observed H{\sc i} morphology provides clear
evidence of the interaction of the SNR with the surrounding
ISM. However, in all cases, the actual distance depends also on the
adopted Galactic rotation curve even in the case where a correlation
is well established. Therefore, the values found is Sect. 5 may be
slightly over (or under) estimated, since some contribution of H{\sc
i} that is not associated with SNR may be included. The kinematic
distances in the second case are in aggrement with previous estimates
of the distance to the SNR, based on several methods summarized in
Section 1. Thus, a distance to the SNR between 6 to 8.5 kpc might be
more plausible. However, more observations are needed (e.g. kinematic)
in order to confidently determine the current stage of evolution of G
32.8$-$0.1.

\section{Conclusions}
Optical emission from the supernova remnant G 32.8$-$0.1 was detected
for a first time with both filamentary and diffuse structure to be
present. The bright optical filaments are correlated very well with
the remnant's radio emission at 1.4 GHz suggesting their association,
while correlation evidences are also shown with the Spitzer Space
Telescope mostly at 8$\mu$m. The flux--calibrated images and the
long--slit spectra indicate that the emission arises from
shock--heated gas. The surrounding ISM was explored through the H{\sc
i} 21 cm emission line. From these observation we adopted a distance
to the SNR between 6 to 8.5 kpc. Finally, an upper limit for the
electron density of 200 \dens\ is calculated.


\begin{acknowledgements}
The authors would like to thank John Meaburn and Jeroen Stil for their
helpful discussions on V$_{LSR}$~calculation. JA acknowledges funding
by the European Union and the Greek Ministry of Development in the
framework of the programme `Promotion of Excellence in Research
Institutes (2nd Part)'. Skinakas Observatory is a collaborative
project of the University of Crete, the Foundation for Research and
Technology-Hellas and the Max-Planck-Institut f\"ur Extraterrestrische
Physik. The National Radio Astronomy Observatory is a facility of the
National Science Foundation operated under cooperative agreement by
Associated Universities, Inc.
\end{acknowledgements}

\newpage
%

%
\begin{table*}  
\caption[]{Imaging and Spectral log}  
\label{table1}
\begin{flushleft} 
\begin{tabular}{lcccccc}  
\noalign{\smallskip}  
\hline  
\multicolumn{6}{c}{IMAGING} \\  
\hline
Filter & $\lambda_{\rm c}$ & $\Delta \lambda$ & Total exp. time &
N$^{\rm o}$ of diff. fields & Telescope\\ & ($\AA$) & ($\AA$) & (sec)
& & \\
\hline
\HNII & 6570 & 75 & 4800 (2)$^{\rm a}$ & 1 & 0.3-m \\
\OIII & 5005 & 28 & 9600 (4) & 1 & 0.3-m \\
\SII & 6720 & 18 & 9600 (4) & 1 & 0.3-m \\
Cont blue & 5470 & 230 & 180 & 1 & 0.3-m \\
Cont red & 6096 & 134 & 180 & 1 & 0.3-m \\
\HNII & 6570 & 75 & 2400 & 6 & 1.3-m \\
Cont red & 6096 & 134 & 180 & 6 & 1.3-m \\
\hline
\multicolumn{7}{c}{SPECTROSCOPY} \\  
\hline  
Area & \multicolumn{2}{c}{Slit centers} & Total exp. time &
Offset$^{\rm b}$ & Aperture length$^{\rm c}$ & Telescope \\  
 & $\alpha$ & $\delta$ & (sec) & (arcsec) & (arcsec)& \\  
\hline  
South I (SI) & 18\h51\m40.6\s & -00\degr18\arcmin04\arcsec & 7800 (2)
& 23.6 N & 13.0 & 1.3-m \\
South II (SII) & 18\h51\m40.6\s & -00\degr18\arcmin04\arcsec  & 7800
 (2) & 8.3 S & 31.9 & 1.3-m\\
East I (EI) & 18\h51\m48.6\s & -00\degr09\arcmin44\arcsec & 3900 &
 37.8 S & 20.1 & 1.3-m\\
East II (EII) & 18\h51\m48.6\s & -00\degr09\arcmin44\arcsec & 3900 & 64.9 S & 10.6 & 1.3-m\\
\hline  
\end{tabular}
\end{flushleft}
\begin{flushleft}
${\rm ^a}$ Numbers in parentheses represent the number of individual frames.\\
${\rm ^b}$ Spatial offset from the slit center in arcsec: N($=$North),
S($=$South).\\ 
${\rm ^c}$ Aperture lengths for each area in arcsec.\\
\end{flushleft} 
\end{table*}

\begin{table}
\caption[]{Typically measured fluxes over the brightest filaments.}
\label{table2}
\begin{flushleft}
\begin{tabular}{lllll}
\hline
\noalign{\smallskip}
 & S & E & N & W \\
\hline
\hnii\  & 20.8 & 19.8 & 18.7 & 21.3 \\
\hline
\sii\   & 17.5 & 14.7 & 12.3 & $<$6$^{\rm a}$\\
\hline
\end{tabular}
\end{flushleft}
Fluxes in units of \flux \\
Median values over a 40\arcsec $\times$ 40\arcsec\ box. \\
S=(South), N(=North), W(=West) and E(=East).\\
${\rm ^a}$ 3$\sigma$~upper limit.\\
 \end{table}

\begin{table*}
\caption[]{Relative line fluxes.}
\label{table3}
\begin{tabular}{lllllllll}
\hline
\noalign{\smallskip}
 & \multicolumn{2}{c}{Area SI} & \multicolumn{2}{c}{Area SII} & \multicolumn{2}{c}{Area EI} & \multicolumn{2}{c}{Area EII}  \\ 
Line (\AA) & F$^{\rm a}$ & S/N$^{\rm b}$ & F & S/N & F & S/N & F & S/N \\
\hline
\hbeta\ 4861 & $<6$ & & $<6$ & & $<4$ & & $<4$ & \\
\oi\ 6300 &  43.5 & (23) & 31.3 & (16) & 31.7 & (33) & 67.0 & (18) \\
\nitrogen\ 6548 & 50.5 & (25) & 54.5 & (25) & 48.6 & (44) & 29.4 & (9)  \\
\ha\  6563 & 100 & (45) & 100 & (45) & 100 & (83) & 100 & (26)  \\
\nitrogen\ 6584 &  167.6 & (64) & 167.1 & (67) & 135.0 & (101) & 98.8 & (25)  \\
\sulfur\ 6716 & 97.1 & (42) & 93.8 & (41) & 88.1 & (70) & 109.8 & (27)  \\
\sulfur\ 6731 & 74.8 & (32) & 73.9 & (32) & 66.4 & (58) & 74.4 & (19)  \\
\hline
Absolute \ha\ flux$^{\rm c}$ & \multicolumn{2}{c}{2.5} & \multicolumn{2}{c}{1.9} & \multicolumn{2}{c}{4.6} & \multicolumn{2}{c}{1.8}  \\
\sulfur/\ha\  & \multicolumn{2}{c}{1.72 $\pm$ 0.05} & \multicolumn{2}{c}{1.68 $\pm$ 0.05} & \multicolumn{2}{c}{1.54 $\pm$ 0.03} & \multicolumn{2}{c}{1.84 $\pm$ 0.09} \\
F(6716)/F(6731) & \multicolumn{2}{c}{1.30 $\pm$ 0.05} & \multicolumn{2}{c}{1.27 $\pm$0.05} & \multicolumn{2}{c}{1.33 $\pm$ 0.03} & \multicolumn{2}{c}{1.48 $\pm$ 0.09} \\
\nii/\ha\ & \multicolumn{2}{c}{2.18 $\pm$ 0.06} &
\multicolumn{2}{c}{2.22 $\pm$ 0.06} & \multicolumn{2}{c}{1.84 $\pm$
0.03}& \multicolumn{2}{c}{1.28 $\pm$ 0.07} \\ 
c(\hbeta)$^{\rm d}$ & \multicolumn{2}{c}{$>$ 2.2} &
\multicolumn{2}{c}{$>$2.2} & \multicolumn{2}{c}{$>$2.7} & \multicolumn{2}{c}{$>$2.7}\\ 
\hline
\end{tabular}

${\rm ^a}$ Observed fluxes normalized to F(H$\alpha$)=100 and uncorrected  
for interstellar extinction. \\ 
${\rm ^b}$ Numbers in parentheses represent the signal to noise ratio of the
quoted fluxes.\\ 
$^{\rm c}$ In units of \flux. \\ 
$^{\rm d}$ The logarithmic extinction is derived by c =
1/0.348$\times$log((\ha/\hbeta)$_{\rm obs}$/2.85).\\
Listed fluxes are a signal to noise weighted average of two fluxes for
area S.\\
The errors of the emission line ratios are calculated through standard
error propagation.\\

\end{table*}

\newpage

\end{document}